\documentclass[review,12pt]{elsarticle}
\usepackage{amssymb}
\usepackage{amsthm}
\usepackage{framed} % To put a box to the nomenclature
\usepackage{amsmath,color}
\usepackage{mathrsfs}
\usepackage{graphicx}
\usepackage{epstopdf}
\usepackage{float}
\usepackage{caption}
\usepackage{subcaption}
\usepackage{bm}
\usepackage{bbm}
\usepackage{mathrsfs}
\usepackage{cleveref}
\usepackage{soul}
\usepackage{accents}
\usepackage{color,soul} %to highlight text
\usepackage{color} %To highlight references in the text
\biboptions{sort&compress}
\usepackage{tabu} % To increase the vertical space between cells in tables
\soulregister\citep7 % To allow citep to be inside of highlighted text
\soulregister\citet7 % Idem with citet
\soulregister\citealp7 % Idem with citealp
\newsavebox{\measurebox} %To create a 1 column figure with a 2 column figure on its side
\usepackage{multirow} %To span several rows and columns
\usepackage{titlesec} %To use paragraph as subsubsubsection
\usepackage[nomarkers,figuresonly]{endfloat} %To place all figures at the end
\usepackage[T1]{fontenc} % This is to convert the paper easily to Word
\usepackage{lmodern} % This is to convert the paper easily to Word
\input{glyphtounicode}  % This is to convert the paper easily to Word

\journal{Journal of Applied Mechanics}

\makeatletter
\def\@author#1{\g@addto@macro\elsauthors{\normalsize%
    \def\baselinestretch{1}%
    \upshape\authorsep#1\unskip\textsuperscript{%
      \ifx\@fnmark\@empty\else\unskip\sep\@fnmark\let\sep=,\fi
      \ifx\@corref\@empty\else\unskip\sep\@corref\let\sep=,\fi
      }%
    \def\authorsep{\unskip,\space}%
    \global\let\@fnmark\@empty
    \global\let\@corref\@empty  %% Added
    \global\let\sep\@empty}%
    \@eadauthor={#1}
}
\makeatother

\titleformat{\paragraph}
{\normalfont\normalsize\itshape}{\theparagraph}{1em}{}
\titlespacing*{\paragraph}
{0pt}{3.25ex plus 1ex minus .2ex}{1.5ex plus .2ex}

\begin{document}

\begin{frontmatter}



\title{Mode II fracture of an elastic-plastic sandwich layer}
%\title{Mode II fracture of an elastic-plastic sandwich layer}
%\title{Mode II fracture of an adhesive jointMode II fracture of an elastic-plastic adhesive layerMode II fracture of an adhesive joint}


\author{Emilio Mart\'{\i}nez-Pa\~neda\fnref{Cam}}

\author{I. Iv\'{a}n Cuesta\fnref{Ubu}}

\author{Norman A. Fleck\corref{cor1} \fnref{Cam}}
\ead{naf1@cam.ac.uk}

\address[Cam]{Department of Engineering, Cambridge University, CB2 1PZ Cambridge, UK}

\address[Ubu]{Universidad de Burgos, Escuela Polit\'{e}cnica Superior, 09006 Burgos, Spain.}

\cortext[cor1]{Corresponding author.}

\begin{abstract}
The shear strength of a pre-cracked  sandwich layer is predicted, assuming that the layer is linear elastic or elastic-plastic, with yielding characterized by either J2 plasticity theory or by a strip-yield model. The substrates are elastic and of dissimilar modulus to that of the layer. Two geometries are analysed: (i) a semi-infinite crack in a sandwich layer, subjected to a remote mode II $K$-field and (ii) a centre-cracked sandwich plate of finite width under remote shear stress. For the semi-infinite crack, the near tip stress field is determined as a function of elastic mismatch, and crack tip plasticity is either prevented (the elastic case) or is duly accounted for (the elastic-plastic case). Analytical and numerical solutions are then obtained for the centre-cracked sandwich plate of finite width. First, a mode II $K$-calibration is obtained for a finite crack in the elastic sandwich layer. Second, the analysis is extended to account for crack tip plasticity via a mode II strip-yield model of finite strength and of finite toughness.  The analytical predictions are verified by finite element simulations and a failure map is constructed in terms of specimen geometry and crack length.

\end{abstract}


\end{frontmatter}



\section{Introduction}
\label{Sec:Introduction}

Multi-material, multi-layer systems are increasingly used in engineering components in order to confer a desired functionality, such as electrical inter-connection, thermal conductivity and mechanical strength. The sensitivity of fracture strength to the presence of defects is a concern, and an appropriate fracture mechanics requires development. In the present study, we consider the idealised case of a compliant layer between two stiffer substrates. Adhesive lap joints are of such a geometry. Adhesively bonded joints can offer significant advantages over competing joining techniques: the advantages include weight reduction, reduced through life maintenance, and fewer sources of stress concentration. Accordingly, there is continued interest in the use of an adhesive layer for bonding applications across the aerospace, transport, energy and marine sectors \cite{Camanho2011,DaSilva2011}. In many of these applications, the adhesive joint is subjected to macroscopic shear loading. However, the shear fracture of adhesives has received only limited attention in the mechanics literature; this motivates the present study. A wide range of constitutive behaviours are shown by adhesive layers, depending upon the material choice. Ceramic  or highly cross-linked polymers behave in an essentially elastic, brittle manner. Soldered and brazed joints comprise a metallic layer, and it is natural to treat these by an elastic-plastic solid. Polymeric adhesives cover an enormous range from  rubber-like behaviour, with high failure strain (at temperatures above the glass transition temperature), to visco-plastic or elastic-brittle (at temperatures below the glass transition temperature). The small strain response can be taken as elastic at temperatures much below the glass transition temperature, to visco-elastic in the vicinity of the glass transition. Thus, it is overly simplistic to treat all polymers at all temperatures as visco-elastic. In the present study, we shall consider the idealised extremes of behaviour of the adhesive layer: it is either treated as elastic-brittle with a finite elastic modulus and finite toughness, or is treated as elastic-ideally plastic, with a finite value of critical crack tip displacement for fracture. The elastic-plastic idealisation is an adequate representation for thermosetting polymers such as toughened epoxy adhesives. More sophisticated choices of adhesive are left to future studies, as our present intent is to explore the role of layer compliance, layer strength and layer toughness upon the macroscopic fracture strength of a layer containing a finite crack. The limiting case of a semi-infinite crack within the layer, and the substrates loaded by a remote mode II K field is also addressed.\\

Insight into the initiation and growth of a mode II crack in an adhesive layer has been gained through tests on End-Notched Flexure (ENF) and Butterfly specimen geometries, see Refs.  \cite{Chai1996,Blackman2005,DaSilva2010,DeMoura2009}, and the references therein. Strip-yield models are used to characterise the fracture response of the adhesive joint, based on an assumed or measured traction-separation law of the adhesive, see, for example, Refs. \cite{Yang2001,Chen2010,Dourado2012,Fernandes2013}.\\

In the present study, we combine theoretical analysis with finite element (FE) modelling to gain insight into the fracture of pre-cracked sandwich layer subjected to macroscopic shear loading. The layer is characterized by linear elasticity, by ideally-plastic, J2 flow theory of plasticity or by a mode II strip-yield model \cite{Dugdale1960}. The substrates are taken to be elastic, and of sufficiently high strength that they do not yield. Two geometries are considered: (i) a boundary layer formulation, whereby a remote $K_{II}$ field is prescribed on a semi-infinite crack within a sandwich layer, and (ii) a centre-cracked plate of finite width, comprising an adhesive layer sandwiched between two elastic substrates, and subjected to a remote shear stress.  The fracture criterion is the attainment of the mode II crack tip toughness:  a critical value of crack tip mode II stress intensity for an elastic strip, or a critical value of crack tip sliding displacement for the strip-yield model or J2 plasticity theory. \\ 

The paper is organized as follows. Section \ref{Sec:BL} contains the analysis of a sandwich layer containing a semi-infinite crack and subjected to a remote mode II $K$-field. First, the layer is treated as elastic but of different modulus to that of the substrates. Then, the analysis is extended to an elastic-plastic layer, with plasticity represented either by a strip-yield model, or by the J2 flow theory of plasticity. Section \ref{Sec:CentreCrackPanel} presents the analytical derivation of the fracture strength of a centre-cracked sandwich panel of finite width, containing a linear elastic layer or an elastic-plastic layer. The mode II $K$-calibration is determined in order to predict the failure strength of an elastic-brittle adhesive layer containing a centre-crack but with no strip-yield zone present. Then, the analysis is extended to account for a crack tip fracture process zone by making use of a mode II strip-yield model of finite strength and of finite toughness. Failure maps are derived for the regimes of behaviour and the analytical predictions are verified by finite element simulations of the strip-yield model. Additional finite element simulations are used for which the layer satisfies J2 flow theory, and the crack tip mode II displacement is compared to that of the strip-yield model. Finally, concluding remarks are given in Section \ref{Sec:Concluding remarks}.

\section{An adhesive layer with a semi-infinite crack}
\label{Sec:BL}

Consider first an elastic layer of thickness $h$ containing a semi-infinite crack and two elastic substrates of modulus that differs from that of the layer. The sandwich plate is subjected to a remote mode II $K$-field of magnitude $K^\infty$, see Fig. \ref{fig:BoundaryLayer}. The crack tip stress state is evaluated for a linear elastic layer in Section \ref{Sec:LinearBL}, and the analysis is then extended to the case of an elastic-plastic layer, with plasticity modelled in Section \ref{Sec:PlasticBL} either by a strip-yield model or by J2 flow theory.

\subsection{An elastic sandwich layer containing a semi-infinite crack}
\label{Sec:LinearBL}

\begin{figure}[H]
  \makebox[\textwidth][c]{\includegraphics[width=1\textwidth]{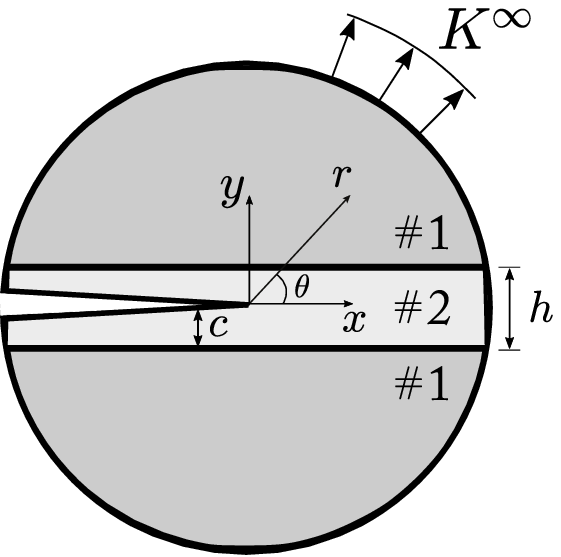}}%
  \caption{Sketch of the boundary layer formulation for an adhesive joint. The substrates are made from material \#1 whereas the adhesive layer is made from material \#2.}
  \label{fig:BoundaryLayer}
\end{figure}

Assume plane strain conditions throughout this study and write $E$ as Young's modulus, $\nu$ as Poisson's ratio, and  $\mu \equiv E /(2 (1+\nu))$ as the shear modulus. As shown in Fig. \ref{fig:BoundaryLayer}, the substrates are made from material 1 (with elastic properties $E_1$, $\nu_1$, and $\mu_1$), and the adhesive layer is made from material 2 (with elastic properties $E_2$, $\nu_2$, and $\mu_2$). We investigate the role of the elastic modulus mismatch between the layer and the substrates. Consider first a crack located at mid-height of the layer, $c/h=0.5$. Then symmetry dictates that the crack tip is in a state of pure mode II. By path-independence of the $J$-integral \cite{Rice1968a}, the remote $K^\infty$ field is related to a local mode II $K^{tip}$ field by
\begin{equation}\label{Eq:KtipKinfty}
K^{tip} = \left[ \frac{E_2 \left( 1 - \nu_1^2 \right)}{E_1 \left( 1 - \nu_2^2 \right)} \right]^{1/2} K^\infty
\end{equation}

Finite element computations of the shear stress distribution $\tau(x)$ at a distance $x$ directly ahead of the crack tip and of the crack tip displacement profile $\delta(x)$ behind the crack tip are conducted for the boundary layer formulation sketched in Fig. \ref{fig:BoundaryLayer}. A remote, elastic mode II $K^\infty$-field is imposed by prescribing a mode II displacement field on the outer periphery of the mesh of the form,
\begin{equation}
u_i = \frac{K^\infty}{\mu_1} r^{1/2} f_i \left( \theta, \, \nu \right)
\end{equation}

\noindent where the functions $f_i \left( \theta, \, \nu \right)$ are written in Cartesian form as \cite{Williams1957}
\begin{equation}
f_x = \sqrt{\frac{2}{\pi}} \left( 2 -2 \nu + \cos^2 \left( \frac{\theta}{2} \right) \right) \sin \left( \frac{\theta}{2} \right)
\end{equation}
\begin{equation}
f_y = - \sqrt{\frac{2}{\pi}} \left( 1 - 2\nu - \sin^2 \left( \frac{\theta}{2} \right) \right) \cos \left( \frac{\theta}{2} \right)
\end{equation}

The finite element model is implemented in the commercial package ABAQUS/Standard \footnote{Abaqus/Standard 2017. Dassault Systemes SIMULIA, Providence, Rhode Island.}. We discretise the geometry by means of fully integrated plane strain, quadratic, quadrilateral elements. Symmetry about the crack plane is exploited when the crack is located at mid-height of the adhesive thickness, such that only the upper half of the domain is analysed; typically, 350,000 degrees-of-freedom are employed.

\subsubsection{Crack tip field: effect of elastic mismatch}

\begin{figure}[H]
  \makebox[\textwidth][c]{\includegraphics[width=0.8\textwidth]{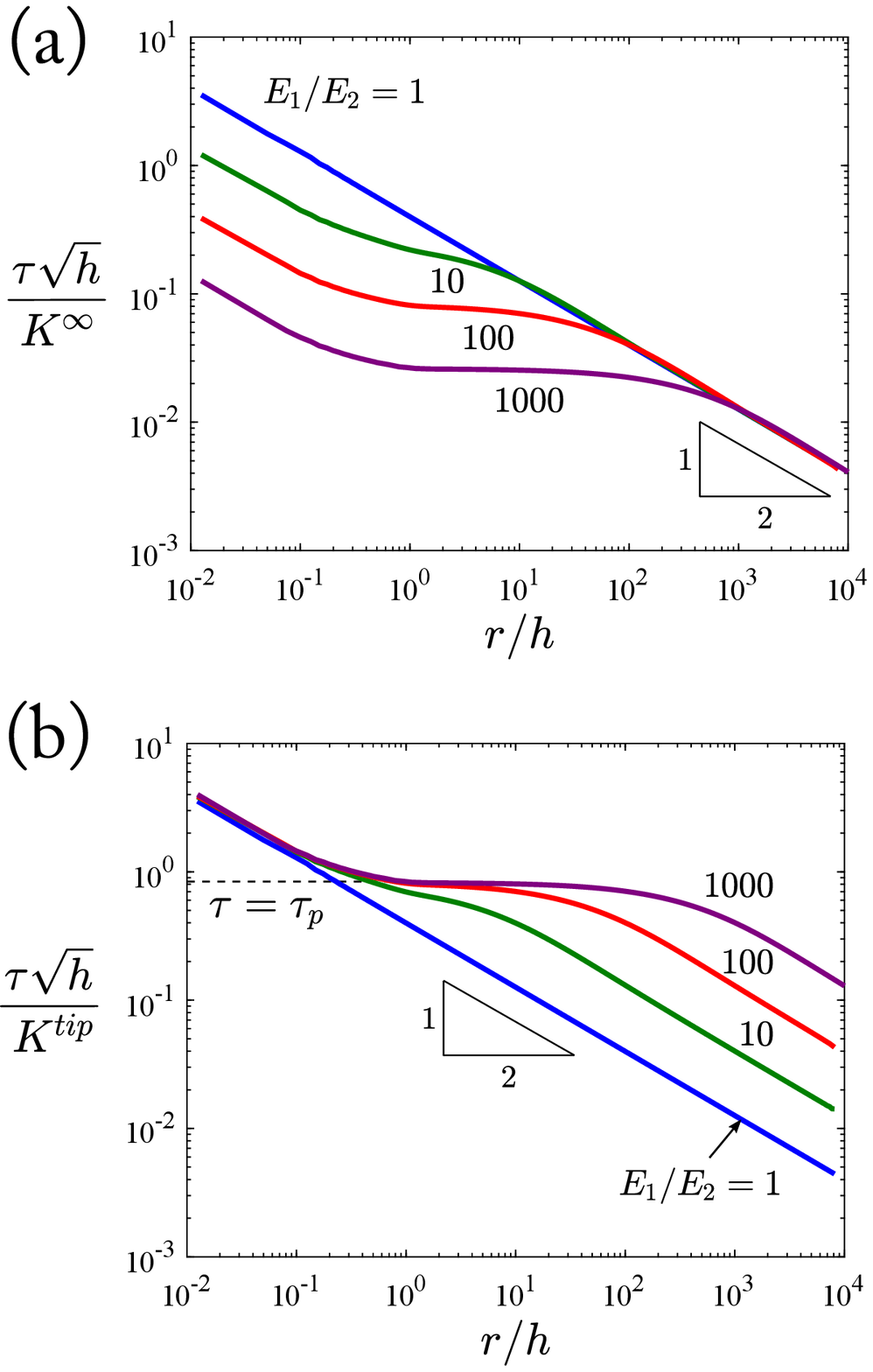}}%
  \caption{Shear stress ahead of a semi-infinite crack in an elastic adhesive, normalized by (a) the remote stress intensity factor $K^\infty$, and (b) the crack tip stress intensity factor $K^{tip}$. Results are shown for selected values of the modulus mismatch $E_1/E_2$, with $\nu=\nu_1=\nu_2=0.3$.}
  \label{fig:Kelastic}
\end{figure}

Consider a semi-infinite crack located at mid-height of the adhesive, as sketched in Fig. \ref{fig:BoundaryLayer}. The finite element prediction for the shear stress distribution $\tau(x)$ directly ahead of the crack tip is shown in Fig. \ref{fig:Kelastic}, for selected values of modulus mismatch $E_1/E_2$ from 1 to 1000; attention is limited, however, to the choice $\nu=\nu_1=\nu_2=0.3$.\\

The shear stress $\tau(x)$ directly ahead of the crack tip is normalised by $K^\infty$ and $K^{tip}$ in Figs. \ref{fig:Kelastic}a and \ref{fig:Kelastic}b, respectively. Both inner and outer $K$-fields exist, and each satisfy the usual $r^{-1/2}$ singularity as analysed by Williams \cite{Williams1957}. Thus, upon making use of the polar coordinate system $(r, \theta)$ centred at the crack tip, the crack tip shear stress distribution in the outer field, along $\theta=0$, is given by
\begin{equation}\label{eq:Williams}
\tau = \frac{K^\infty}{\sqrt{2 \pi r}} = \frac{K^{tip}}{\sqrt{2 \pi r}} \left[ \frac{E_1 \left( 1 - \nu_2^2 \right)}{E_2 \left( 1 - \nu_1^2 \right)} \right]^{1/2}
\end{equation}
Likewise, the inner field is of the form 
\begin{equation}\label{eq:Wiinner}
\tau = \frac{K^{tip}}{\sqrt{2 \pi r}}
\end{equation}

\noindent Note from Fig. \ref{fig:Kelastic}b that the inner and outer $K$-fields are connected by a region of almost constant shear stress $\tau_p$ of magnitude $\tau_p \sqrt{h}/K^{tip} \approx 1$. The extent of this zone enlarges with increasing modulus mismatch between layer and substrate.\\

The relation $\tau_p \sqrt{h}/K^{tip} \approx 1$ between plateau stress $\tau_p$ ahead of the crack tip and $K^{tip}$ agrees with the following  analytical result for an elastic strip of modulus $E_2$ and Poisson ratio $\nu_2$ sandwiched between two rigid substrates and subjected to a remote shear stress $\tau_p$. Following Rice \cite{Rice1968a}, the upstream work density of the sandwich layer of height $h$ is given by
\begin{equation}\label{eq:UpstreamWork}
W^U=\frac{1}{2} \frac{\tau_p^2 h}{\mu_2}
\end{equation}

\noindent per unit area of layer. Upon noting that the downstream work density vanishes, the energy release rate is $G=W^U$. Now make use of the usual Irwin relation between $G$ and the mode II crack tip stress intensity factor $K^{tip}$ such that
\begin{equation}
\left( K^{tip} \right)^2 =  \frac{E_2}{\left(1-\nu_2^2 \right)} G = \frac{E_2}{2 \left(1-\nu_2^2 \right) \mu_2} \tau_p^2 h 
\end{equation}

\noindent It follows immediately that
\begin{equation}\label{eq:plateau}
\frac{\tau_p \sqrt{h}}{K^{tip}} =(1-\nu_2)^{1/2}
\end{equation}

\begin{figure}[H]
  \makebox[\textwidth][c]{\includegraphics[width=1\textwidth]{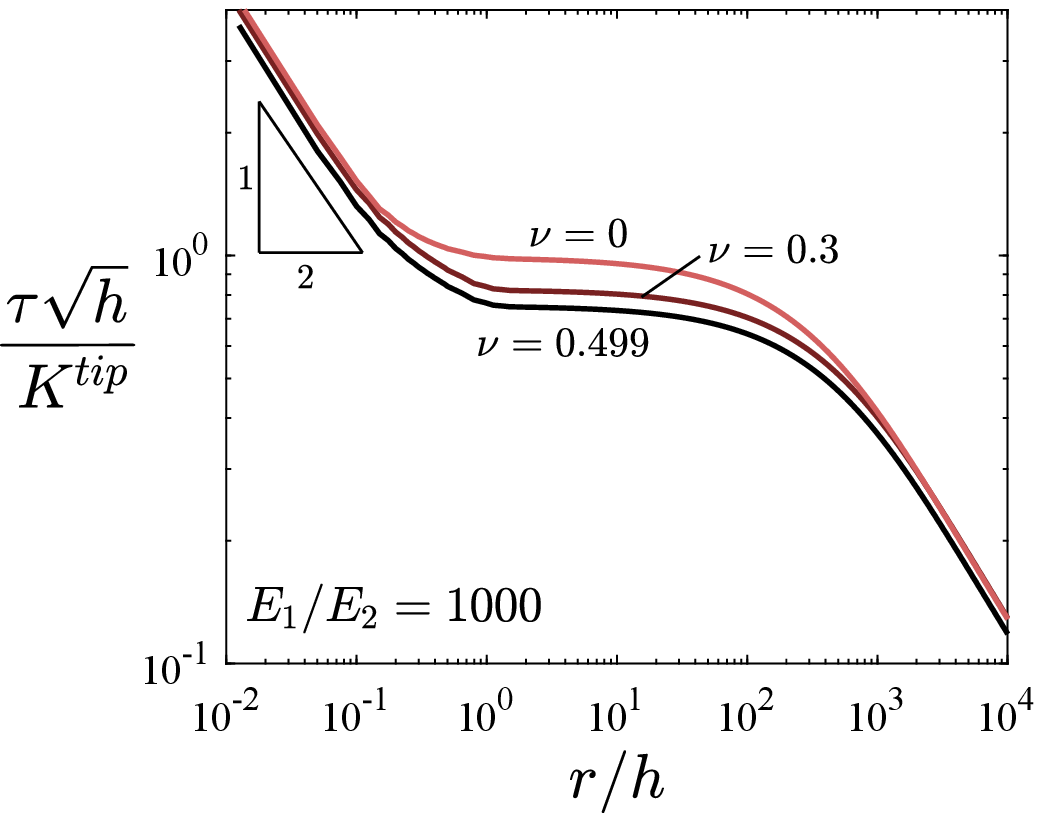}}%
  \caption{Influence of the Poisson's ratio ($\nu=\nu_1=\nu_2$) on the shear stress ahead of a semi-infinite crack in an elastic adhesive. The shear stress is normalized by the crack tip stress intensity factor $K^{tip}$.}
  \label{fig:Ktipnu}
\end{figure}

\noindent Thus, the magnitude of the plateau shear stress $\tau_p \sqrt{h}/K^{tip}$ depends only upon the Poisson's ratio of the adhesive layer in the limit $E_2/E_1 \to 0$. The sensitivity of the stress distribution to Poisson's ratio is investigated numerically in Fig. \ref{fig:Ktipnu} for $E_1/E_2=1000$. The plateau stress $\tau_p$ increases slightly with decreasing $\nu$, and the predictions of Eq. (\ref{eq:plateau}) are in good agreement with the numerical predictions.

The boundary between the zone of dominance of the plateau stress and that of the outer remote $K$-field occurs at a distance $r=\lambda$ from the crack tip. The magnitude of $\lambda$ is estimated by equating the values of shear stress in (\ref{eq:Williams}) and (\ref{eq:plateau}) at $r=\lambda$, to give
\begin{equation}
\frac{\lambda}{h} = \frac{1}{2 \pi } \frac{E_1 \left( 1 + \nu_2 \right)}{E_2 \left( 1 - \nu_1^2 \right)}
\end{equation}

Thus, for the choice $E_1/E_2=1000$ and $\nu=0.3$, the plateau stress region extends a distance of $\lambda/h=227$ ahead of the crack tip; the finite element results agree with this estimation, see Fig. \ref{fig:Kelastic}b. This large value of $\lambda/h$ has an immediate practical implication: the required crack length and in-plane structural dimensions in order for a remote $K$ field to exist is on the order of  meters for a polymeric adhesive layer of height $h=5$mm sandwiched between metallic or ceramic substrates. This puts a severe limitation on the applicability of a conventional fracture mechanics assessment of the fracture strength of a polymer-based adhesive layer sandwiched between substrates of much higher modulus.\\

\subsubsection{Mixed mode ratio: influence of crack location and elastic properties}

Consider now the influence of the crack location with respect to the height of the adhesive layer upon the mode mix. The plane of the crack is quantified by the parameter $c/h$, with $c/h=0.5$ denoting a crack at mid-height and $c/h=0$ denoting a crack on the lower interface between the strip and the substrate. As noted by Dundurs \cite{Dundurs1969} (see also, Hutchinson and Suo \cite{Hutchinson1991}), a wide class of plane problems in isotropic elasticity of bimaterial interfaces can be formulated  in terms of only two material parameters: $\alpha$ and $\beta$. For the case of plane strain, the Dundur's parameters read
\begin{equation}
\alpha= \frac{\mu_2 \left(1 - \nu_1 \right) - \mu_1 \left(1 - \nu_2 \right)}{\mu_2 \left( 1 - \nu_1 \right) + \mu_1 \left( 1 - \nu_2 \right)}
\end{equation}
\begin{equation}
\beta=\frac{1}{2} \frac{\mu_2 \left(1 - 2\nu_1 \right) - \mu_1 \left(1 - 2 \nu_2 \right)}{\mu_2 \left( 1 - \nu_1 \right) + \mu_1 \left(1 - \nu_2 \right) }
\end{equation}

\noindent Thus, $\beta$ vanishes when both materials are incompressible ($\nu_1=\nu_2=0.5$). The values of $\alpha$ and $\beta$ corresponding to the elastic properties assumed throughout this work are listed in Table \ref{tab:Dundurs}.

\begin{table}[H]
\centering
\caption{Dundurs' parameters for the values of $\nu$ and $E_1/E_2$ adopted.}
\label{tab:Dundurs}
   {\tabulinesep=2mm
   \begin{tabu}{|c | c c | c c | c c | c c |} 
\hline
$\nu$ & \multicolumn{2}{|c|}{$E_1/E_2=3$} & \multicolumn{2}{|c|}{$E_1/E_2=10$} & \multicolumn{2}{|c|}{$E_1/E_2=100$} & \multicolumn{2}{|c|}{$E_1/E_2=1000$} \\
\hline
   & \multicolumn{1}{|c}{$\alpha$} & \multicolumn{1}{|c|}{$\beta$} & \multicolumn{1}{|c}{$\alpha$} & \multicolumn{1}{|c|}{$\beta$} & \multicolumn{1}{|c}{$\alpha$} & \multicolumn{1}{|c|}{$\beta$} & \multicolumn{1}{|c}{$\alpha$} & \multicolumn{1}{|c|}{$\beta$} \\ \cline{2-9}
0 & \multicolumn{1}{|c|}{0.500} & 0.250 & \multicolumn{1}{|c|}{0.818} & 0.409 & \multicolumn{1}{|c|}{0.980} & 0.490 & \multicolumn{1}{|c|}{0.998} & 0.499 \\
0.3 & \multicolumn{1}{|c|}{0.500} & 0.140 & \multicolumn{1}{|c|}{0.818} & 0.234 & \multicolumn{1}{|c|}{0.980} & 0.280 & \multicolumn{1}{|c|}{0.998} & 0.285 \\
0.49 & \multicolumn{1}{|c|}{0.500} & 0.010 & \multicolumn{1}{|c|}{0.818} & 0.016 & \multicolumn{1}{|c|}{0.980} & 0.019 & \multicolumn{1}{|c|}{0.998} & 0.020 \\ \hline
   \end{tabu}}
\end{table}

\begin{figure}[H]
  \makebox[\textwidth][c]{\includegraphics[width=0.85\textwidth]{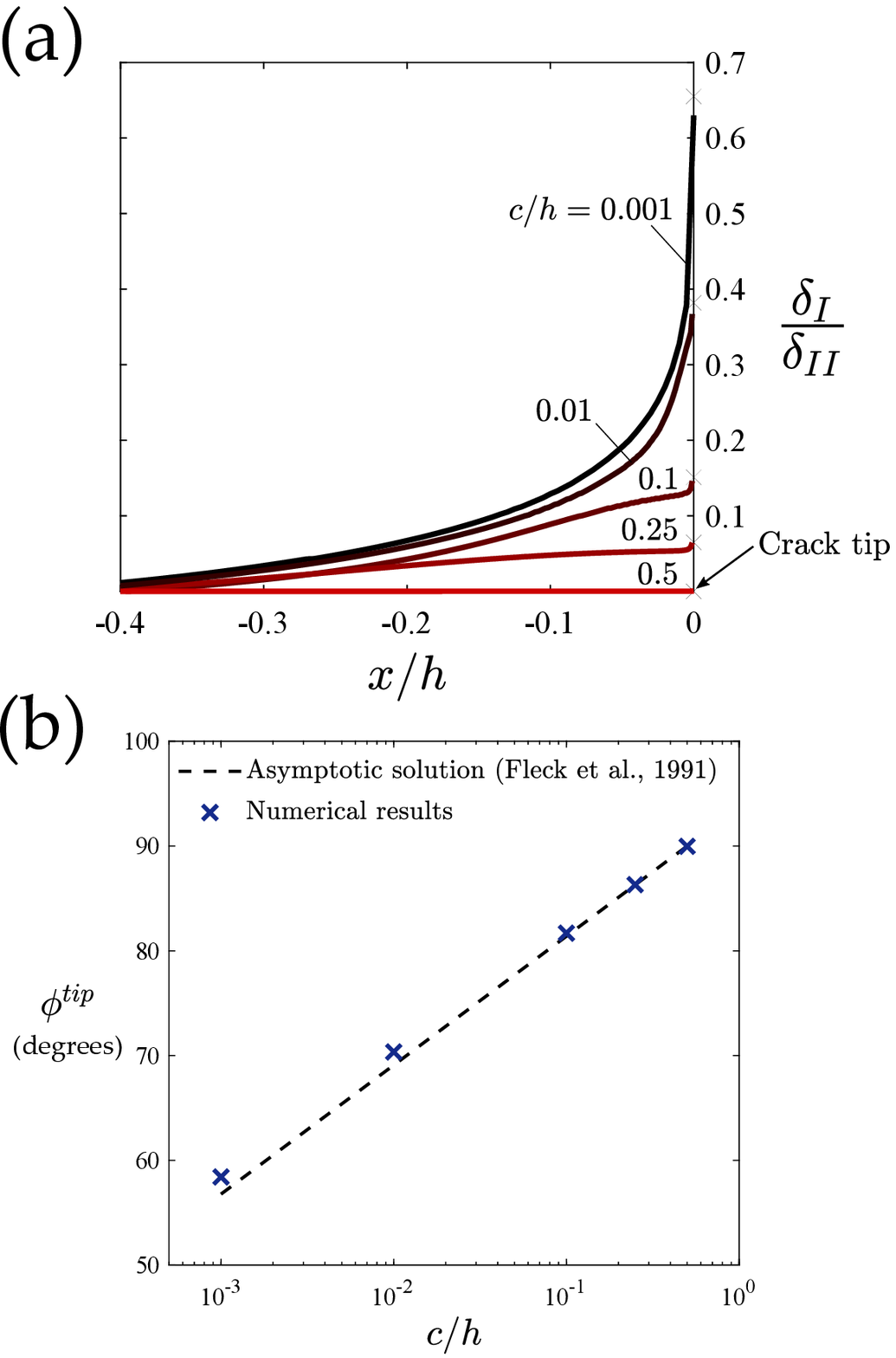}}%
  \caption{Mode mix as a function of the crack plane in relation to  the height of the layer: (a) crack tip displacement ratio, and (b) phase angle at the crack tip. Elastic modulus mismatch $E_1/E_2=1000$, Poisson's ratio: $\nu_1=\nu_2=0.3$.}
  \label{fig:MixedModeCrackLocation}
\end{figure}

In the present study, the mode mix in the vicinity of the crack tip is characterized in terms of the relative opening displacement $\delta_I$ to sliding displacement  $\delta_{II}$ behind the crack tip. Consider first the sensitivity of mode mix to $c/h$, for the choice $E_1/E_2=1000$ and $\nu_1=\nu_2=0.3$. Finite element predictions are shown in Fig. \ref{fig:MixedModeCrackLocation}a. Remote mode II loading leads to mixed mode loading at the crack tip ($x=0$), with the ratio $\delta_I/\delta_{II}$ increasing as the crack plane approaches the interface, $c/h \to 0$. The presence of the finite mode II stress intensity at the crack tip implies that the crack will tend to kink into the interface. For all values of $c/h$ considered, the magnitude of $\delta_I/\delta_{II}$ drops sharply with increasing distance from the crack tip, with the local mode I contribution becoming negligible for $|x|>0.3h$. \\

The sensitivity of mode mix at the crack tip to the value of $c/h$ has been analysed previously by Fleck et al. \cite{Fleck1991}. They obtained the asymptotic behaviour of the crack tip phase angle $\phi=\tan^{-1} (\delta_{II} / \delta_{I})$. They showed that $\phi$ depends upon the crack plane location with regard to the layer height $c/h$ and to the Dundur's parameters $\alpha$ and $\beta$ according to their equation (10) and restated here as
\begin{equation}
\phi = \epsilon \ln \left(\frac{h-c}{c}\right) + 2 \left( \frac{c}{h} - \frac{1}{2} \right) \left( \phi_H \left( \alpha, \, \beta  \right) + \omega \left( \alpha, \, \beta \right) \right)
\end{equation}

\noindent where 
\begin{equation}
\epsilon=\frac{1}{2 \pi} \ln \left(\frac{1-\beta}{1+\beta}\right)
\end{equation}
The functions $\phi_H \left( \alpha, \, \beta  \right) $ and $\omega \left( \alpha, \, \beta \right)$ have been tabulated previously in Refs. \cite{Hutchinson1987,Suo1989}. The
 numerically computed values of the crack tip phase angle $\phi$ are compared with the asymptotic solution of Fleck et al. \cite{Fleck1991} in Fig. \ref{fig:MixedModeCrackLocation}b; excellent agreement is observed, in support of the accuracy of the finite element simulations of the present study.  \\

\begin{figure}[H]
  \makebox[\textwidth][c]{\includegraphics[width=1.4\textwidth]{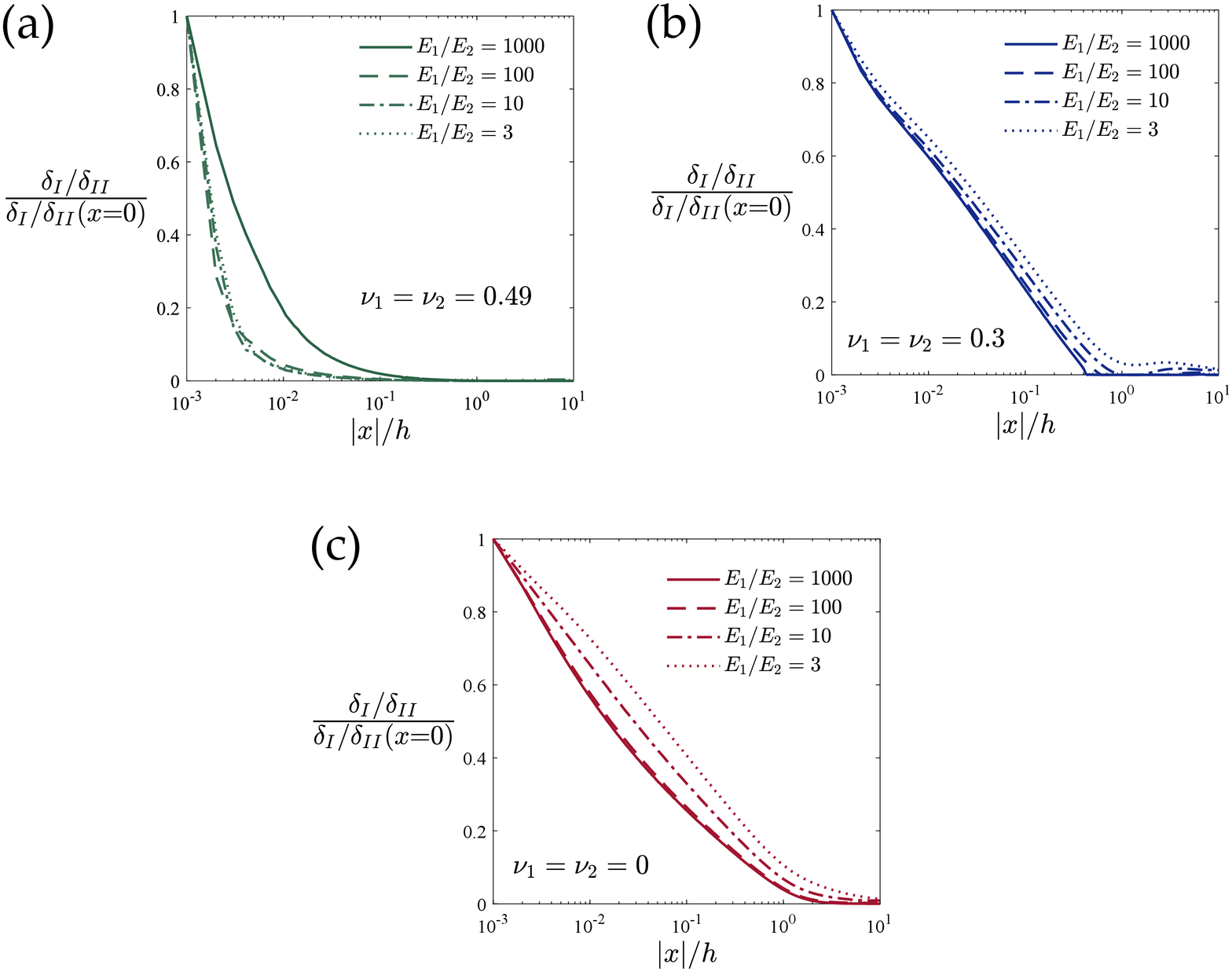}}%
  \caption{Ratio of opening to sliding displacements behind crack tip, for selected values of the modulus mismatch. Results are shown for  selected values of Poisson's ratio $\nu=\nu_1=\nu_2$: (a) $\nu=0.49$, (b) $\nu=0.3$, and (c) $\nu=0$. The crack plane is located at $c/h=0.001$.}
  \label{fig:MixedModeSeparationLog}
\end{figure}

We proceed to investigate the effect of material mismatch $E_1/E_2$ and Poisson's ratio $\nu=\nu_1=\nu_2$ upon the near tip displacement field for a crack that lies very close to the lower interface, $c/h=0.001$. The mode mix, $\delta_I / \delta_{II}$, normalized by the mode mix at $x=0$ is plotted as a function of distance $x/h$ behind the crack tip in Fig. \ref{fig:MixedModeSeparationLog}; for completeness, the numerical values obtained for $\delta_I / \delta_{II} (x=0)$ are given in Table \ref{tab:Separation0}.

\begin{table}[H]
\centering
\caption{Numerically computed values of mixed mode ratio of crack tip displacement $\delta_I/\delta_{II}$ at the crack tip ($x=0$) as a function of the values of $\nu$ and $E_1/E_2$. The crack plane is located at $c/h=0.001$}
\label{tab:Separation0}
   {\tabulinesep=2mm
   \begin{tabu}{|c c c c c|} 
\hline
 \multicolumn{5}{|c|}{$\delta_I/\delta_{II} \left( x=0 \right)$} \\
\hline
 $\nu$  & \multicolumn{1}{|c|}{$E_1/E_2=3$} & $E_1/E_2=10$ & \multicolumn{1}{|c|}{$E_1/E_2=100$} & $E_1/E_2=1000$ \\ \hline
0  & \multicolumn{1}{|c|}{0.59} & 1.21 & \multicolumn{1}{|c|}{1.84} & 1.93 \\
0.3  & \multicolumn{1}{|c|}{0.29} & 0.48 & \multicolumn{1}{|c|}{0.55} & 0.62 \\
0.49  & \multicolumn{1}{|c|}{0.06} & 0.10 & \multicolumn{1}{|c|}{0.09} & 0.02 \\  \hline
   \end{tabu}}
\end{table}

\noindent The finite element results, as presented in Figs. \ref{fig:MixedModeSeparationLog}a-c reveal only a  small influence of modulus mismatch and of  Poisson's ratio upon the normalised mode mix, unless $\nu$ is close to the incompressible limit of $\nu=0.5$. The ratio of crack opening to crack sliding displacement is significant only close to the crack tip; this domain decreases from approximately $h$ to  $0.01h$ (with the precise value depending upon the modulus mismatch) as $\nu$ approaches 0.5.  These results justify the choice of a pure mode II strip-yield model for the analysis of crack growth in adhesive joints subjected to remote mode II $K^\infty$ loading, provided that the strip-yield zone is of length $h$ or greater. 

\subsection{Elastic-plastic adhesive with a semi-infinite crack}
\label{Sec:PlasticBL}

Consider now the influence of plastic deformation upon the crack tip stress and strain state in the sandwich layer by assuming that the layer behaves as an elastic, ideally plastic von Mises solid.

\subsubsection{Influence of plasticity on crack tip mode mix}

\begin{figure}[H]
  \makebox[\textwidth][c]{\includegraphics[width=0.75\textwidth]{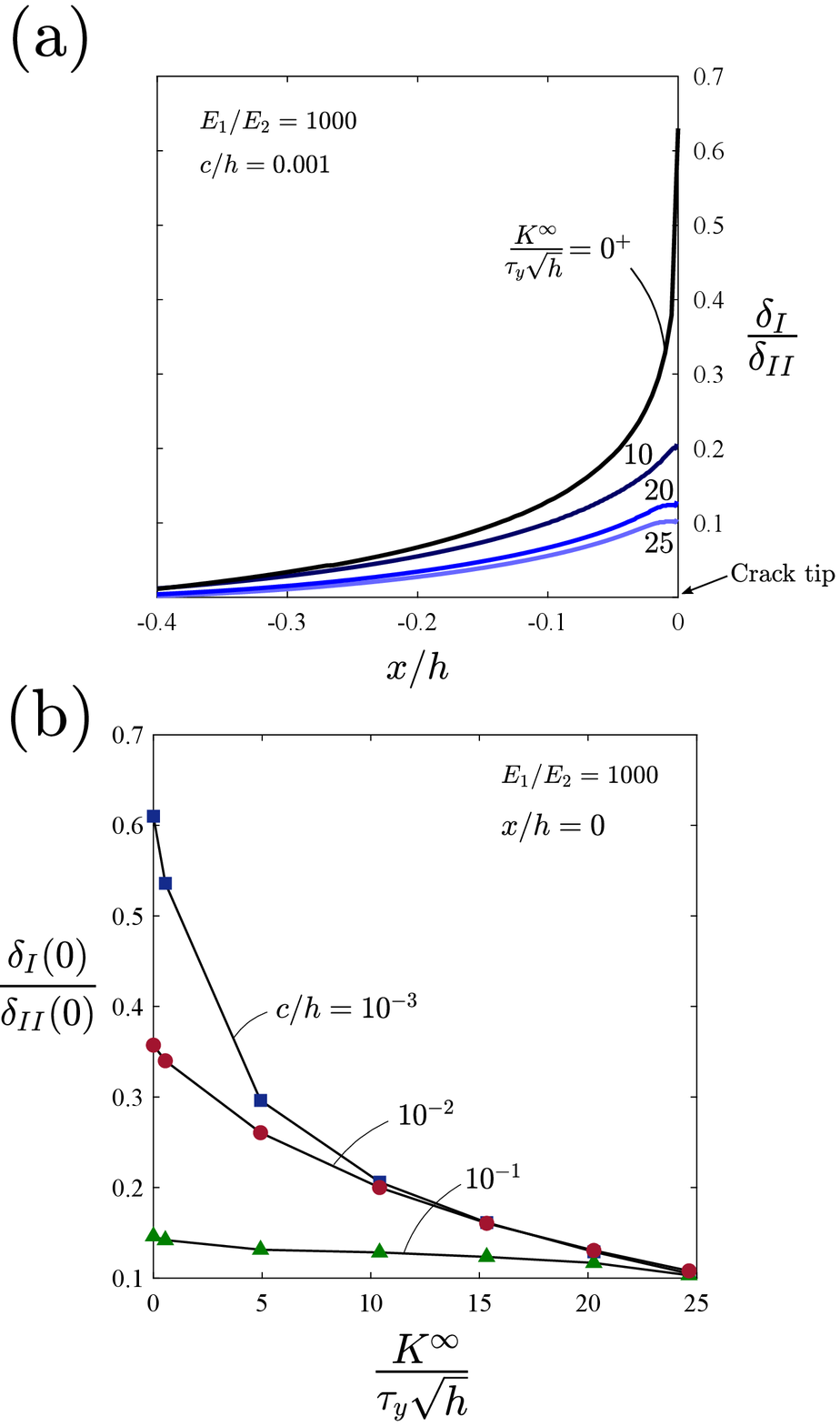}}%
  \caption{ (a) Ratio of opening to sliding displacements versus distance from crack tip, for selected values of remote K-value, with plasticity present. (b) Ratio of opening to sliding displacement at the crack tip.}
  \label{fig:MixedModedelta}
\end{figure}

First, we assess the role of plasticity in influencing the tensile and shear crack tip displacements. Thus, we conduct similar calculations to those reported in Section \ref{Sec:LinearBL} but with the sandwich layer now characterized by J2 plasticity theory, for the choice $\tau_y/\mu_1=6.5 \times 10^{-6}$. (Note that the plastic zone size, and the mode mix are insensitive to the value of this parameter, whereas the crack tip displacement is sensitive to its value.)  The distribution of mode ratio $\delta_I/\delta_{II}$ behind the crack tip is shown in Fig. \ref{fig:MixedModedelta}. Results are presented for selected values of load intensity $K^\infty/(\tau_y \sqrt{h})$. The dominance of mode II over mode I displacements increases with the degree of plasticity and with increasing $c/h$ (up to 0.5, for which $\delta_I=0$), see Fig. \ref{fig:MixedModedelta}b. These results strengthen the conclusions of the previous section: crack tip plasticity ensures that the crack tip is close to mode II in nature provided the remote loading is mode II.

\subsubsection{Strip-yield model to represent crack tip plasticity}
\label{Sec:DugdalevsJ2plasticity}

\begin{figure}[H]
  \makebox[\textwidth][c]{\includegraphics[width=0.8\textwidth]{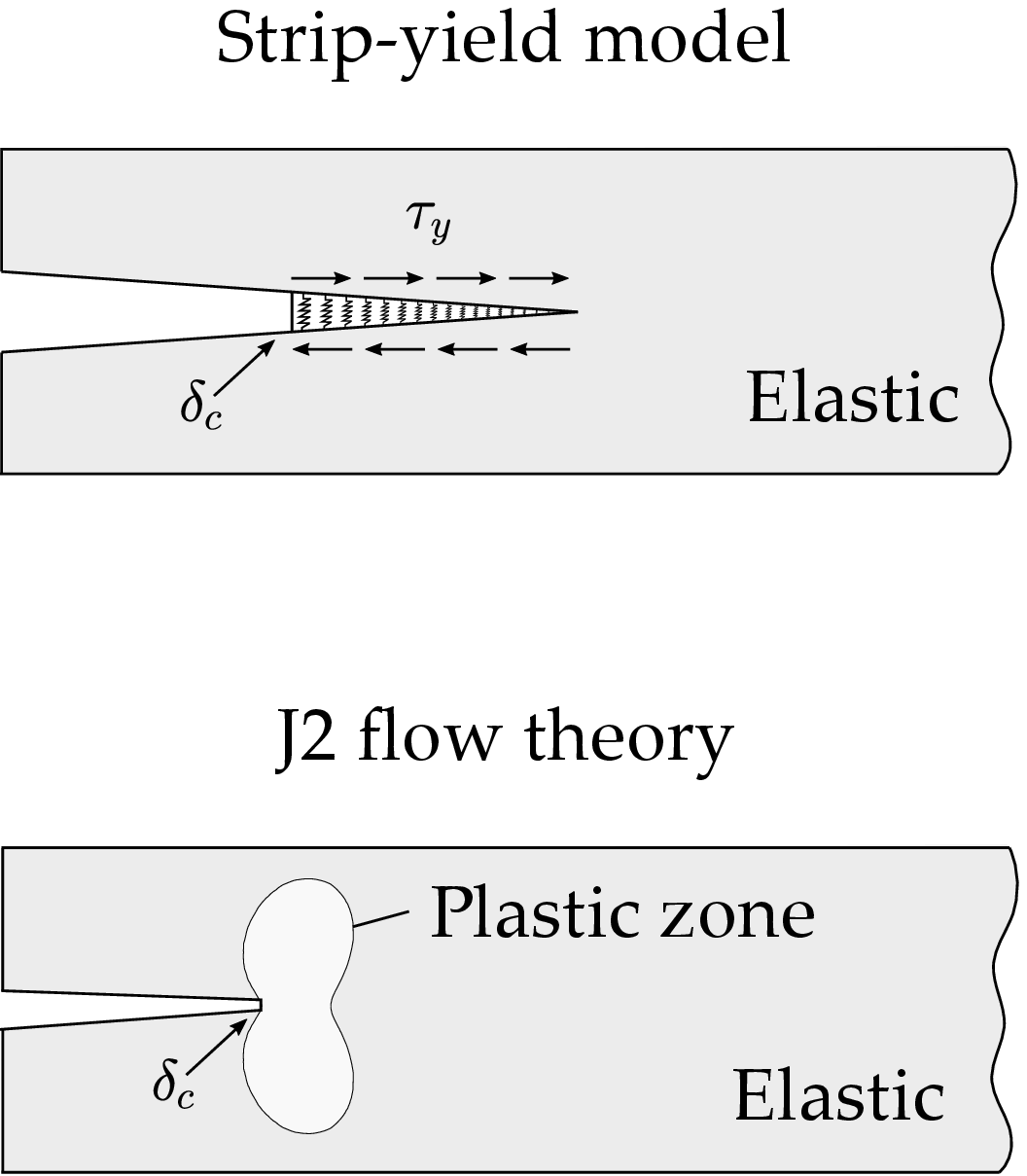}}%
  \caption{Sketch of the strip-yield and J2 plasticity theory idealisations  for modelling plasticity within the layer.}
  \label{fig:DugdaleSketch}
\end{figure}

We shall now show that the strip-yield model provides a good approximation to the plastic zone size as obtained for J2-flow theory. Specifically, we employ the shear yield version of Dugdale's strip-yield model \citep{Dugdale1960}. The traction-separation law is characterised by a finite shear strength $\tau_y$. The strip-yield model is implemented in ABAQUS/Standard by making use of cohesive elements, see Ref. \cite{EFM2017} for details. In brief, mode I opening is suppressed within the cohesive zone by a penalty function, and only mode II sliding along the cohesive zone surfaces is permitted.  A total of approximately 20,000 plane strain, quadratic elements with full integration have been used, with the same mesh employed for the strip-yield calculation and for the case of J2-flow theory (absent a cohesive zone). A sketch of both approaches is given in Fig. \ref{fig:DugdaleSketch}.\\

\begin{figure}[H]
  \makebox[\textwidth][c]{\includegraphics[width=1\textwidth]{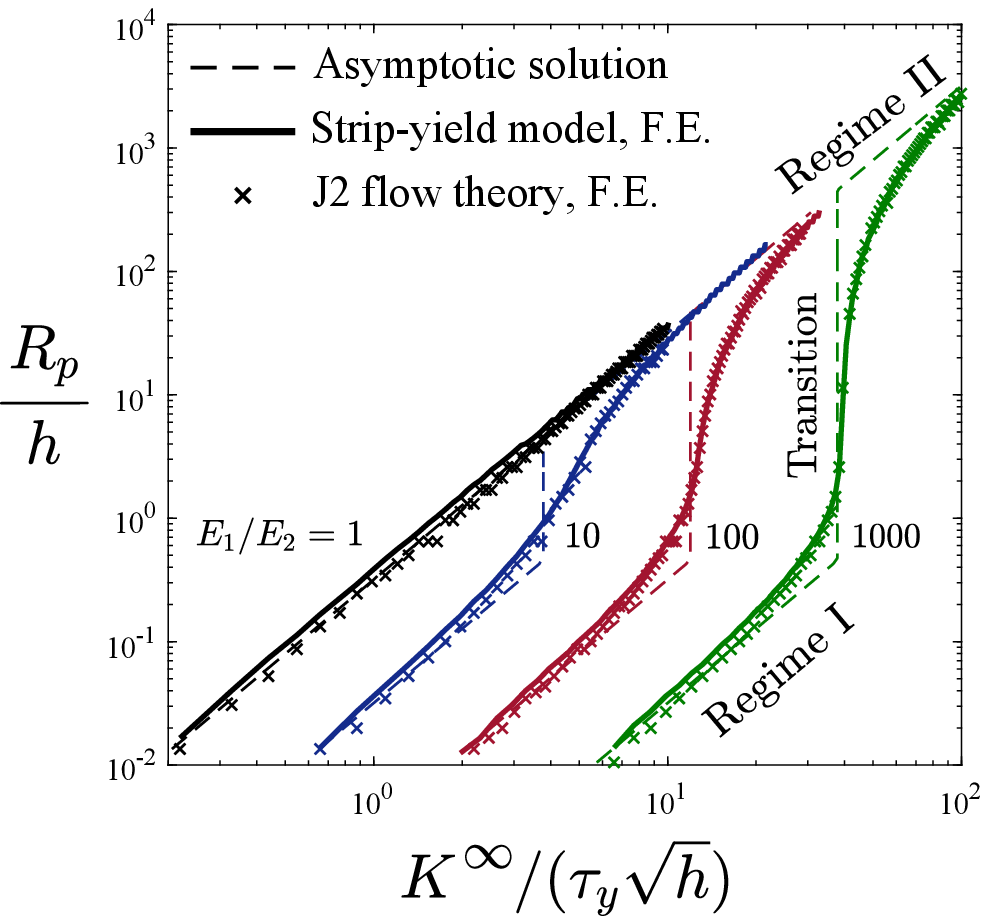}}%
  \caption{Plastic zone size as a function of the remote stress intensity factor for strip-yield model and for J2 flow theory, for selected values of modulus mismatch.}
  \label{fig:RpvsKtipDugdale}
\end{figure}

Finite element predictions of the plastic zone size $R_p$ are shown in Fig. \ref{fig:RpvsKtipDugdale} as a function of remote stress intensity for selected values of Young's modulus mismatch: $E_1/E_2=1$, $10$, $100$ and $1000$. The numerical predictions obtained with J2 plasticity theory and the strip-yield model approximation are in excellent agreement.\\

Two distinct regimes can be identified: regime I, as given by
\begin{equation} \label{eq:RegimeI}
R_p = \frac{1}{\pi} \left( \frac{K^{tip}}{\tau_y} \right)^2
\end{equation}

\noindent and regime II, as given by
\begin{equation}\label{eq:RegimeII}
R_p = \frac{1}{\pi} \left( \frac{K^\infty}{\tau_y} \right)^2
\end{equation}

\noindent These regimes are shown by dashed lines in Fig. \ref{fig:RpvsKtipDugdale} and the asymptotic behaviours are supported by the finite element predictions. Note that $R_p$ is independent of the modulus mismatch in regime II but is sensitive to $E_1/E_2$ in regime I.\\

It is clear from Fig. \ref{fig:RpvsKtipDugdale} that the transition from Regime I to Regime II occurs at a transition value of $K^\infty / \left( \tau_y \sqrt{h} \right)$ that scales linearly with $\left( E_1/E_2 \right)^{1/2}$. In other words, the transition value occurs when $K^{tip}/(\tau_y \sqrt{h})$ attains a specific value, upon noting the identity (\ref{Eq:KtipKinfty}). This transition is explained as follows.\\

Recall the trajectory of the shear stress  $\tau (r)$ versus distance $r$ ahead of the crack tip for the case of an elastic layer, as summarised in Fig. \ref{fig:Kelastic}b. With increasing distance $r$ from the crack tip, $\tau(r)$ scales as $\tau=K^{tip}/\sqrt{2 \pi r}$, then $\tau$ equals $\tau_p$, as given in (\ref{eq:plateau}), and then $\tau$ scales as $\tau=K^{\infty}/\sqrt{2 \pi r}$.\\

In regime I, the crack tip plastic zone resides within the $K^{tip}$-field and $\tau_y>\tau_p$, implying via (\ref{eq:plateau}) that
\begin{equation}\label{eq:RegimeIKtip}
K^{tip} < \frac{\tau_y \sqrt{h}}{\left( 1 - \nu_2 \right)^{1/2}}
\end{equation}

This criterion, when re-phrased in terms of an inequality of $h$,
\begin{equation}\label{eq:hinequality}
h > \left( 1 - \nu_2^2 \right) \left( \frac{K^{tip}}{\tau_y} \right)^2
\end{equation}

\noindent is in good agreement with the usual ASTM size criterion \cite{ASTM1820} for the existence of a crack tip $K$-field in the presence of crack tip plasticity,
\begin{equation}\label{eq:ASTMh}
h > 2.5 \left( \frac{K^{tip}}{\tau_y} \right)^2
\end{equation}

\noindent upon taking $h$ to be the leading structural dimension. The small difference in the constants contained within (\ref{eq:hinequality}) and (\ref{eq:ASTMh}) is noted, but does not imply an inconsistency within the analysis: (\ref{eq:ASTMh}) is slightly more restrictive than (\ref{eq:hinequality}). Now make use of (\ref{eq:RegimeI}) to re-write (\ref{eq:RegimeIKtip}) in the form
\begin{equation}
R_p < \frac{1}{\pi} \frac{h}{\left( 1 - \nu_2 \right)}
\end{equation}

\noindent thereby confirming the interpretation that Regime I exists when the plastic zone size $R_p$ is smaller than the layer thickness.\\

Now consider regime II. It pre-supposes that the plastic zone $R_p$ resides within the outer $K$-field, such that $\tau_y<\tau_p$ in Fig. \ref{fig:Kelastic}b. This inequality can be re-written in terms of $K^\infty$ via (\ref{Eq:KtipKinfty}) and (\ref{eq:plateau}) as
\begin{equation}
\frac{K^\infty}{\tau \sqrt{h}} > \frac{1}{\left( 1-\nu_2\right)^{1/2}}\left( \frac{E_1}{E_2} \right)^{1/2} 
\end{equation}

\noindent This transition value of $K^\infty/(\tau_y \sqrt{h})$ is in good agreement with the finite element predictions of Fig. \ref{fig:RpvsKtipDugdale}.\\

The large jump in value of $R_p$ at the transition from regime I to regime II (see Fig. \ref{fig:RpvsKtipDugdale}) is associated with the jump in value of the plastic zone size as determined by the intersection point of the horizontal line $\tau=\tau_y$  and the $\tau (x)$ curve of Fig. \ref{fig:Kelastic}b. As $\tau_y$ is decreased from a value above $\tau_p$ to a value below $\tau_p$ there is a discontinuous jump in the intersection point.\\

\begin{figure}[H]
  \makebox[\textwidth][c]{\includegraphics[width=1\textwidth]{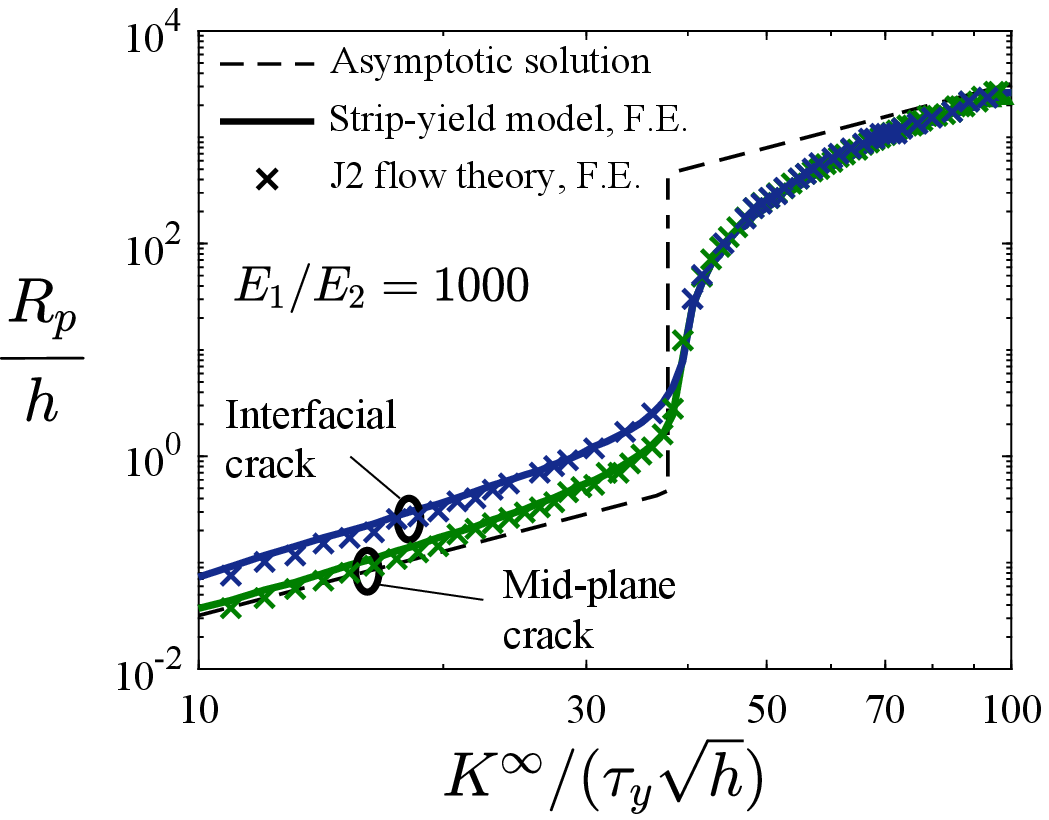}}%
  \caption{Plastic zone size as a function of the stress intensity factor at the crack tip and the crack location, according to the strip-yield model and according to J2 flow theory.}
  \label{fig:DugdalevsBulkCrackLocationKinfty}
\end{figure}

\begin{figure}[H]
  \makebox[\textwidth][c]{\includegraphics[width=1.2\textwidth]{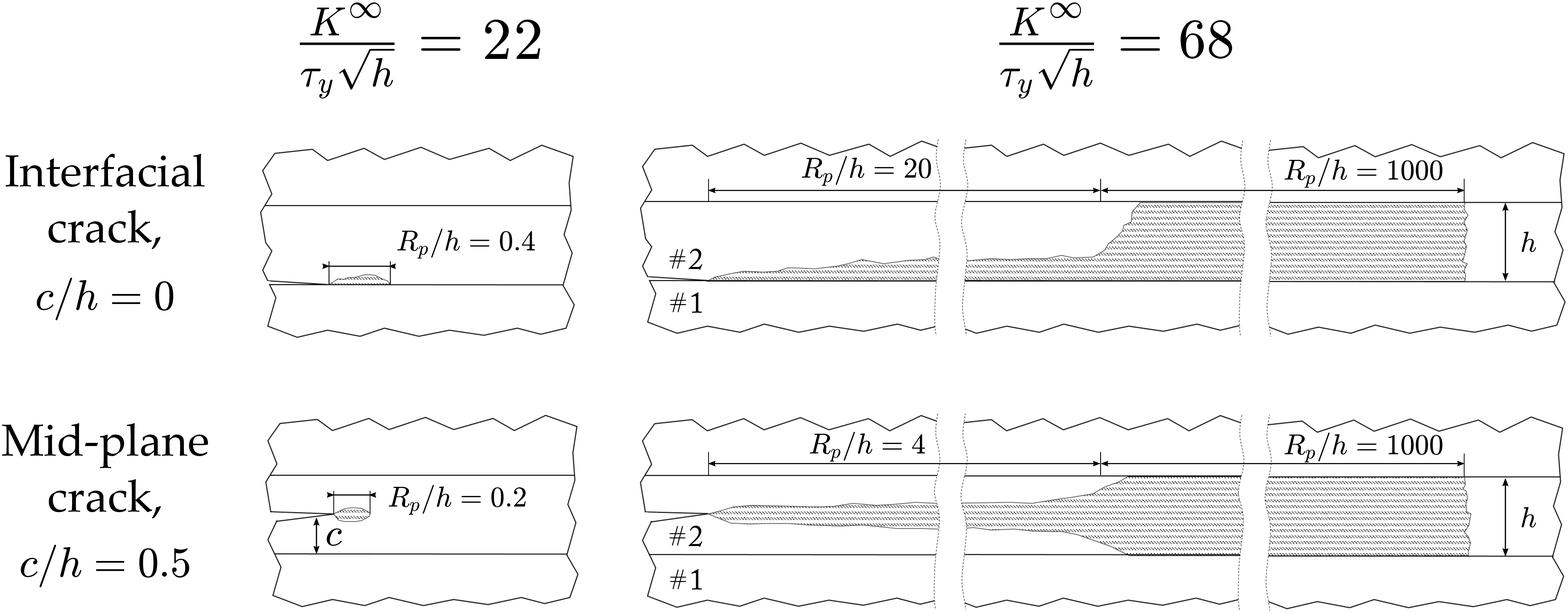}}%
  \caption{The crack tip plastic zone for a crack located in the mid-plane and along the lower interface. The left side shows results obtained at $K^{\infty}=22 \tau_y \sqrt{h}$ while the right side shows results obtained at $K^{\infty}=68 \tau_y \sqrt{h}$.}
  \label{fig:PlasticZoneContours}
\end{figure}

Consider now the case where the crack is not located on the mid-plane but resides along the upper or lower interface of the layer. The dependence of the plastic zone size upon $K^{\infty}$ is shown in Fig. \ref{fig:DugdalevsBulkCrackLocationKinfty} for $E_1/E_2=1000$. The predictions for upper or lower interfacial cracks are identical, as dictated by symmetry. However, interfacial cracks have larger plastic zones  than mid-plane cracks at low remote loads (regime I). In regime II, the size of the plastic zone is independent of the location of the crack. The shape of the plastic zones is shown in Fig. \ref{fig:PlasticZoneContours} for a crack at mid-height of the sandwich layer, and for a crack along the lower interface.  In all cases, the strip-yield model gives an excellent approximation to the plastic zone size as predicted by J2 flow theory.\\

\section{Fracture strength of a centre-cracked adhesive joint}
\label{Sec:CentreCrackPanel}

We proceed to explore the strength of a centred-cracked sandwich plate subjected to a remote shear stress $\tau^\infty$. Consider a sandwich layer of height $h$ and width $2W$, sandwiched between two substrates, and containing a centre crack of length $2a$. We first derive in Section \ref{Sec:PlateElastic} an analytical solution for the fracture strength by assuming that the sandwich layer is linear elastic, and  then extend the analysis to the elastic-plastic case in Section \ref{Sec:PlatePlastic} by means of a strip-yield model of fracture energy $\Gamma$ and cohesive strength $\tau_y$.
We emphasise that the cohesive shear strength is taken to equal the shear yield strength. This is a consequence of the elastic, ideally plastic assumption for the bulk behaviour of the adhesive. This assumption also finds experimental support: commonly, the measured value of fracture strength of polymeric adhesives is comparable to their yield strength \cite{Blackman2003a,Salomonsson2008,Sun2008,Sun2008a,Carlberger2010,Stigh2010} The analysis extends the recent work of Van Loock et al. \cite{VanLoock2019} from mode I fracture of a centre-cracked sandwich layer to the mode II case.\\
 
It is recognised that, in general, layer toughness may not only depend upon $h$ but also upon the degree of crack extension if the adhesive joint exhibits significant crack growth resistance. However, a negligible R-curve is observed for thin, polymer-based adhesive joints; see  Tvergaard and Hutchinson \cite{Tvergaard1994,Tvergaard1996} and Van Loock et al. \cite{VanLoock2019}. \\

\subsection{Crack in an elastic layer}
\label{Sec:PlateElastic}

Write the compliance $C$ of a centre-cracked sandwich plate in terms of the shear displacement $u$ and load $P$, such that $C=u/P$. Then, the extra compliance due to the presence of the crack of length $2a$ is $\Delta C \left( a \right) = C \left( a \right) - C \left( 0 \right)$, and the energy release rate for crack advance $G$ is given by \cite{Tada2000}
\begin{equation}\label{eq:defG}
G = \frac{P^2}{4} \frac{\partial \left( \Delta C \right)}{\partial a}
\end{equation}

\begin{figure}[H]
  \makebox[\textwidth][c]{\includegraphics[width=1.2\textwidth]{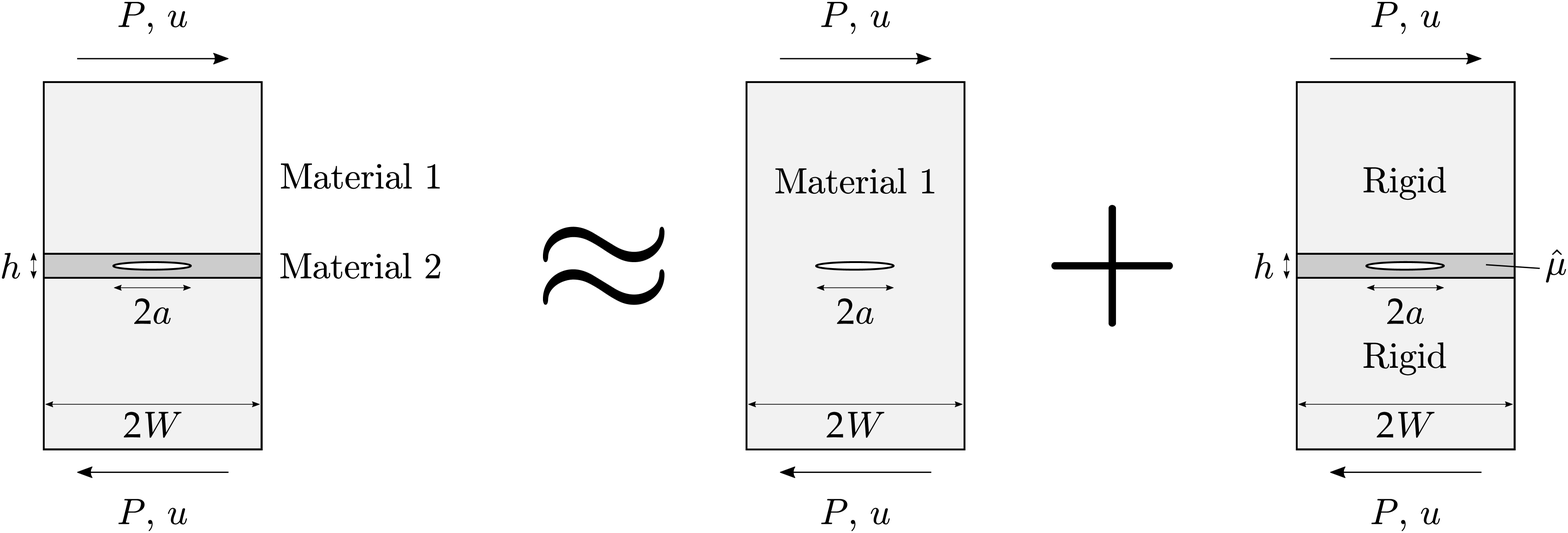}}%
  \caption{Sketch of the superposition procedure employed to calculate the macroscopic shear strength of a centre-cracked sandwich layer.}
  \label{fig:Superposition}
\end{figure}

We proceed to use the superposition principle and idealise the adhesive joint system by the summation of the two problems, as depicted in Fig. \ref{fig:Superposition}: (1) a homogeneous plate with the elastic properties of the substrates, and (2) an adhesive joint with shear modulus $\hat{\mu}$ constrained between two rigid substrates. Accordingly, the variation of the compliance reads,
\begin{equation}
\frac{\partial \left( \Delta C \right)}{\partial a} = \frac{\partial \left( \Delta C^{(1)} \right)}{\partial a} + \frac{\partial \left( \Delta C^{(2)} \right)}{\partial a}
\end{equation}

We seek expressions for $\partial \left( \Delta C^{(1)} \right) / \partial a$ and $\partial \left( \Delta C^{(2)} \right) / \partial a$. As shown in Ref. \cite{Tada2000}, $\partial \left( \Delta C^{(1)} \right) / \partial a$ is given by 
\begin{equation}\label{Eq:DevComp1}
\frac{\partial \left( \Delta C^{(1)} \right)}{ \partial a} = \frac{\pi a F^2 \left(1 - \nu_1^2 \right)}{W^2 E_1}
\end{equation}

\noindent with the finite width correction factor being \cite{Tada2000}
\begin{equation}\label{eq:F}
F = \left[ 1 - 0.025 \left( \frac{a}{W} \right)^2 + 0.06 \left( \frac{a}{W} \right)^4 \right] \left[ \sec \left( \frac{\pi}{2} \frac{a}{W} \right) \right]^{1/2}
\end{equation}

For the second problem, the extra compliance due to the presence of the crack, $\Delta C (a)$, can be readily derived as
\begin{equation}
\Delta C^{(2)}\left( a \right) = \frac{h}{2W \hat{\mu} (1-a/W)}
\end{equation}

\noindent and consequently,
\begin{equation}\label{Eq:DevComp2}
\frac{\partial \Delta C^{(2)}}{\partial a}  = \frac{h}{2W^2 \hat{\mu} (1 - a/W)^2}
\end{equation}

\noindent where the shear modulus of the adhesive $\hat{\mu}$ is given by
\begin{equation}
\frac{1}{\hat{\mu}} = \frac{1}{\mu_2} - \frac{1}{\mu_1}
\end{equation}

Considering Eq. (\ref{eq:defG}) and making use of the Irwin relationship, $K^{tip}=\sqrt{E_2 G/(1-\nu_2^2)}$, the crack tip stress intensity factor (assumed mode II) is given by
\begin{equation}\label{eq:Ktip}
K^{tip} = \frac{P}{2} \left( \frac{E_2}{(1-\nu^2_2)} \frac{\partial \left( \Delta C \right)}{\partial a} \right)^{1/2}
\end{equation}

We now introduce the normalized shear strength as
\begin{equation}\label{eq:shearstrength}
\bar{\tau} = \frac{\tau^\infty_f \sqrt{h}}{\sqrt{E_2 \Gamma /(1-\nu_2^2)}} = \frac{P}{2W} \frac{\sqrt{h}}{K^{tip}}
\end{equation}

\noindent Finally, we substitute Eqs. (\ref{Eq:DevComp1}), (\ref{Eq:DevComp2}), and (\ref{eq:Ktip}) into Eq. (\ref{eq:shearstrength}) in order to obtain a general formula for the strength of an adhesive joint with a centre crack subjected to shear loading:
\begin{equation}\label{eq:tauGriffith1}
\bar{\tau} = \left[ \frac{E_2 \left( 1 - \nu_1^2 \right)}{E_1 \left( 1 - \nu_2^2 \right)} \frac{a}{h} \pi F^2 + \frac{1}{\left( 1 - \nu_2 \right)} \left(1 - \frac{\mu_2}{\mu_1} \right) \left( 1 - \frac{a}{W} \right)^{-2} \right]^{-1/2}
\end{equation}

This general result can be simplified by assuming $\nu_1=\nu_2$ to give
\begin{equation}\label{eq:tauGriffith2}
\bar{\tau} = \left[ \frac{E_2}{E_1} \frac{a}{h} \pi F^2 + \frac{1}{\left( 1 - \nu_2 \right)} \left(1 - \frac{\mu_2}{\mu_1} \right) \left( 1 - \frac{a}{W} \right)^{-2} \right]^{-1/2}
\end{equation}

\noindent and, consistent with Eq. (\ref{eq:plateau}), the limiting case where $a << W$ and $\mu_2 << \mu_1$ reads
\begin{equation}
\bar{\tau} = \frac{\tau \sqrt{h}}{K^{tip}} =\left(1- \nu_2 \right)^{1/2}
\end{equation}

\begin{figure}[H]
  \makebox[\textwidth][c]{\includegraphics[width=1.2\textwidth]{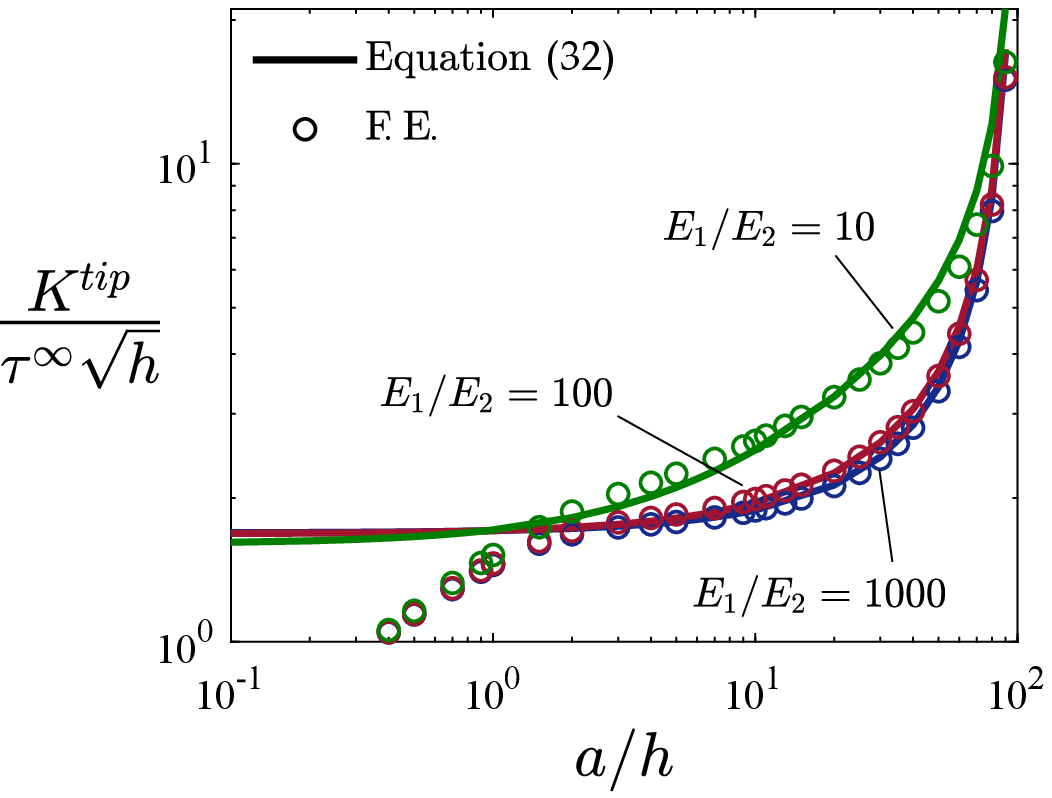}}%
  \caption{$K$-calibration for a crack in an elastic sandwich layer between elastic substrates. $h/W=0.01$}
  \label{fig:Kcallibration}
\end{figure}

The accuracy of equation (\ref{eq:tauGriffith2}) is verified by computing the crack tip stress intensity factor using a finite element formulation with $h/W=0.01$ and selected values of modulus mismatch $E_1/E_2=10,100,1000$. The model is implemented in the commercial finite element package ABAQUS, employing a total of approximately 15000 quadratic quadrilateral elements with full integration. The mode II stress intensity factor $K^{tip}$ is computed by means of an interaction integral method. Results are shown in Fig. \ref{fig:Kcallibration}. Excellent agreement is observed for a crack semi-length $a$ exceeding the layer thickness $h$. 

\subsection{Strip-yield model for a crack in an elastic-plastic layer}
\label{Sec:PlatePlastic}

We now consider a centre-cracked sandwich plate containing an elastic-plastic layer. Assume that fracture occurs at a critical value of the mode II displacement at the crack tip; analytical solutions for the fracture strength of the sandwich plate are derived and the resulting failure maps are displayed to give the strength as a function of geometry and crack length.  The analytical predictions are based upon a strip yield model and the accuracy of these predictions is subsequently verified by a finite element analysis of the same strip yield model.\\

 Consider again the centre-cracked sandwich layer as sketched on the left-hand side of Fig. \ref{fig:Superposition}. We idealize crack tip plasticity by the strip-yield model \cite{Dugdale1960}, as characterized by a rectangular traction-separation law of shear strength $\tau_y$ and toughness $\Gamma$. The critical separation $\delta_c$ follows immediately from the relation $\Gamma=\tau_y \delta_c$. It is  convenient to introduce a reference length scale $l_s$ in the form \cite{JAM2018}
\begin{equation}\label{eq:ls}
l_s = \frac{1}{\pi(1-\nu_2^2)}\frac{E_2 \Gamma}{\tau_y^2} 
\end{equation}

\noindent  which has the interpretation of a representative plastic zone size at the onset of fracture for a long crack. By making use of the strategy of Van Loock et al. \cite{VanLoock2019} (who considered the mode I case) our analysis of the mode II problem is split into three parts. First, we derive the analytical solution for the case of a short crack (Section \ref{Sec:ShortCrack}). Then we consider an intermediate and long crack length (Section \ref{Sec:IntermediateCracks}). And finally, we construct a failure maps to identify competing regimes of behaviour, Section \ref{Sec:Regimes}, and verify the analytical results by a series of finite element calculations (Section \ref{Sec:NumVerify}).

\subsubsection{Short cracks}
\label{Sec:ShortCrack}

Consider first the case where the crack length $a$ is much smaller than both the material length scale $l_s$ and the layer thickness $h$. Then, the strength of the adhesive joint can be predicted by ignoring the presence of the substrate. Accordingly, one can then make direct use of Dugdale's approximation for the crack tip displacement, as given by
\begin{equation}\label{eq:deltaTIP}
\delta^{tip} = \frac{8 \tau_ya (1-\nu_2^2)}{\pi E_2} \ln \left[ \sec \left( \frac{\pi \tau^\infty}{2 \tau_y} \right) \right]
\end{equation}

Now, $\delta_c$ is the value of $\delta^{tip}$ at $\tau^\infty = \tau^\infty_f$. Then, we can re-write (\ref{eq:deltaTIP}) in the form
\begin{equation}\label{eq:shortcrack}
\frac{l_s}{h} = \frac{8}{\pi^2} \frac{a}{h} \ln \left[ \sec \left( \frac{\pi \tau^\infty}{2 \tau_y} \right) \right]
\end{equation}

\noindent  for the characteristic length of the process zone if $a+l_s <<h$ and $h/W \leq 1$. 
%\begin{equation}
%\frac{l_s}{h} = \frac{E_2 \delta_c}{\pi \tau_y h (1-\nu_2^2)} = \frac{8}{\pi^2} \frac{a}{h} \ln \left[ \sec \left( \frac{\pi \tau^\infty}{2 \tau_y} \right) \right]
%\end{equation}

\subsubsection{Intermediate crack lengths}
\label{Sec:IntermediateCracks}

Now suppose that the crack on the order of, or longer than, $l_s$. As in Section \ref{Sec:PlateElastic}, we suppose that the crack tip sliding displacement is the sum of the displacements in the two problems as depicted on the right side of Fig. \ref{fig:Superposition},
\begin{equation}
\delta^{tip} = \delta^{(1)} + \delta^{(2)}
\end{equation}

We first determine $\delta^{(1)}$. The crack tip sliding displacement for a crack of length $2a$ in a linear elastic solid, and subjected to a remote shear stress $\tau^\infty$, is given by
\begin{equation}
\delta^{(1)} = \frac{8 \tau_y a (1-\nu_2^2)}{\pi E_2} \ln \left[ \sec \left( \frac{\pi \tau^\infty}{2 \tau_y} \right) \right]
\end{equation}

In contrast, we deduce the crack tip sliding displacement for the second problem, $\delta^{(2)}$, from the value of the $J$-integral at the crack tip,
\begin{equation}
J^{tip} = \tau_y \delta^{(2)}
\end{equation}

Path-independence of the J-integral implies that $J^{tip}$ equals the value of the $J$-integral taken around a remote contour, $J^\infty$. In addition, $J^\infty$ equals the energy release rate, $G$, which can be deduced from the derivative of the compliance (\ref{Eq:DevComp2}). Accordingly,
\begin{equation}
J^\infty = \frac{h \left( \tau^\infty \right)^2}{2 \hat{\mu}} \left( 1 - \frac{a}{W} \right)^{-2}
\end{equation}

The crack tip sliding displacement, by superimposition of the solution to problems (1) and (2), reads
\begin{equation}
\frac{\delta^{tip}}{h} = \frac{8 \tau_y (1-\nu_1^2)}{\pi E_1} \frac{a}{h} \ln \left[ \sec \left( \frac{\pi \tau^\infty}{2 \tau_y} \right) \right] + \frac{\left( \tau^\infty \right)^2}{2 \hat{\mu} \tau_y} \left( 1 - \frac{a}{W} \right)^{-2}
\end{equation}

Now recall the definition of $l_s$, Eq. (\ref{eq:ls}), and the definition of the fracture energy: $\Gamma=\tau_y \delta_c$. At fracture, $\tau^\infty = \tau^\infty_f$ and $\delta^{tip}=\delta_c$, thereby giving
\begin{equation}\label{Eq:lsFINAL}
\frac{l_s}{h} = \frac{8}{\pi^2} \frac{E_2 \left( 1 - \nu_1^2 \right)}{E_1 \left( 1 - \nu_2^2 \right)} \frac{a}{h} \ln \left[ \sec \left( \frac{\pi \tau^\infty_f}{2 \tau_y} \right) \right] + \frac{ 1}{\pi (1 - \nu_2)} \left( \frac{\tau^{\infty}_f}{\tau_y} \right)^2 \left( 1 -\frac{\mu_2}{\mu_1} \right) \left( 1 - \frac{a}{W} \right)^{-2}
\end{equation}

For the choice $\nu=\nu_1=\nu_2$, this general result simplifies to
\begin{equation}\label{Eq:lsFINAL2}
\frac{l_s}{h} = \frac{8}{\pi^2} \frac{E_2}{E_1} \frac{a}{h} \ln \left[ \sec \left( \frac{\pi \tau^\infty_f}{2 \tau_y} \right) \right] + \frac{ 1}{\pi (1 - \nu)} \left( \frac{\tau^{\infty}_f}{\tau_y} \right)^2 \left( 1 -\frac{\mu_2}{\mu_1} \right) \left( 1 - \frac{a}{W} \right)^{-2}
\end{equation}

Both (\ref{Eq:lsFINAL2}) and (\ref{eq:tauGriffith2}) lead to very similar predictions for $\tau^\infty < \tau_y$ and small $a/W$ values. In fact, one can readily show that both equations predict almost identical results in the limit of $\tau^\infty /\tau_y \to 0$. In this limit  (\ref{Eq:lsFINAL2}) has the asymptotic form
\begin{equation}\label{eq:PlasticTau}
\frac{1}{\bar{\tau}^2} = \pi \frac{E_2}{E_1} \frac{a}{h} + \frac{1}{(1 - \nu)} \left( 1 -\frac{\mu_2}{\mu_1} \right) \left( 1 - \frac{a}{W} \right)^{-2}
\end{equation}

\noindent while (\ref{eq:tauGriffith2}) reduces to
\begin{equation}\label{eq:ElasticTau}
\frac{1}{\bar{\tau}^2} = \pi \frac{E_2}{E_1} \frac{a}{h} F^2 + \frac{1}{(1 - \nu)} \left( 1 -\frac{\mu_2}{\mu_1} \right) \left( 1 - \frac{a}{W} \right)^{-2}
\end{equation}

Thus, the only difference is the presence of the finite width correction factor $F$ in the first term on the right hand side of (\ref{eq:PlasticTau}). As evident from (\ref{eq:F}), $F \approx 1$ for small values of $a/W$.

\subsubsection{Failure map: regimes of behaviour}
\label{Sec:Regimes}

\begin{figure}[H]
  \makebox[\textwidth][c]{\includegraphics[width=0.59\textwidth]{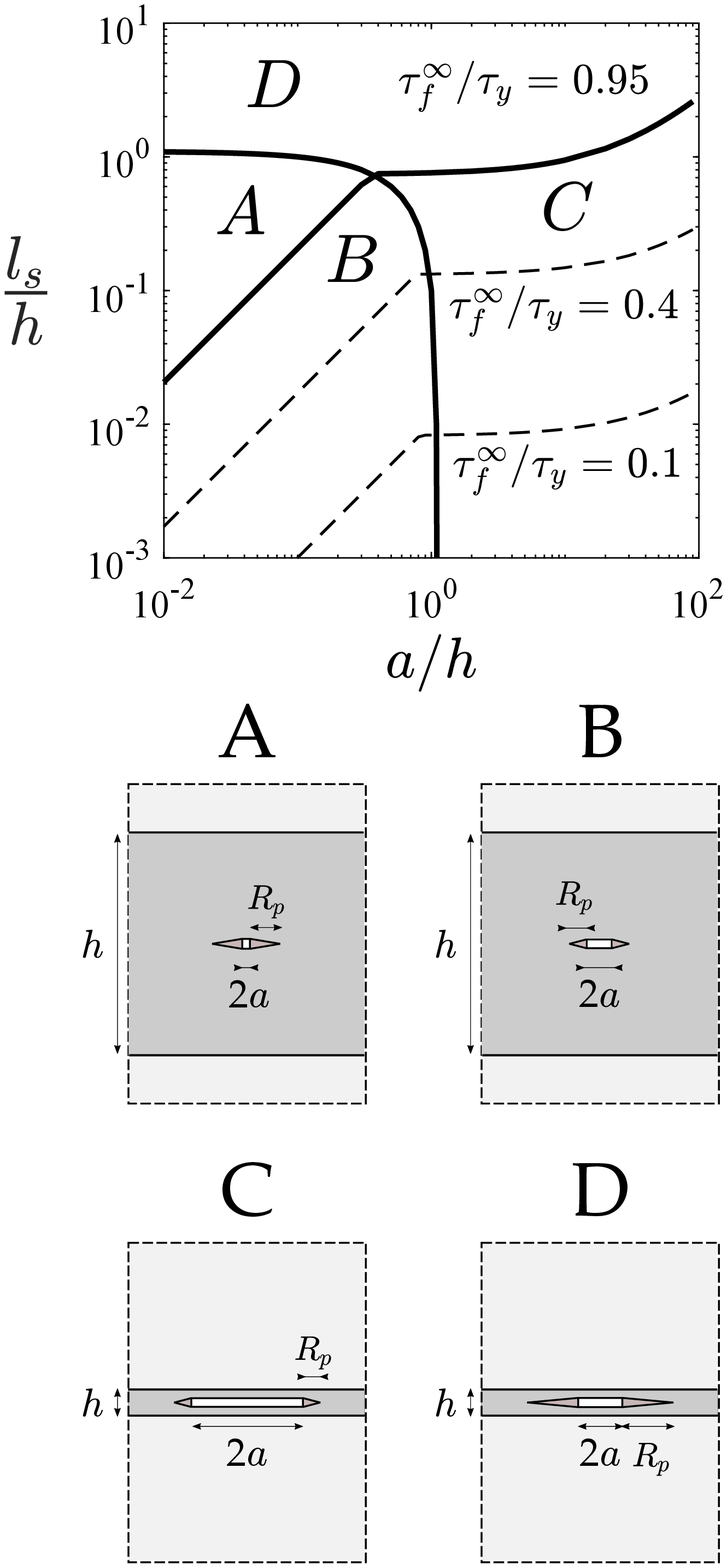}}%
  \caption{Failure map of $l_s/h$ versus normalized crack length $a/h$ for $h/W \to 0$ and mismatch $E_1/E_2=100$. The contours of strength are given by (\ref{eq:tauGriffith2}) in regimes A and B, and by (\ref{Eq:lsFINAL2}) in regimes C and D.}
  \label{fig:MapFigure}
\end{figure}

Upon making use of equations (\ref{eq:shortcrack}) and (\ref{Eq:lsFINAL2}), failure maps can be constructed in terms of specimen geometry and crack length, see Fig. \ref{fig:MapFigure} for the choice $h/W \to 0$ and $E_1/E_2=100$.  Three selected values of $\tau^\infty_f/\tau_y$ are assumed: 0.1, 0.4 and 0.95. The choice $\tau^\infty_f/\tau_y=0.95$ defines the boundary between cohesive zone toughness-controlled fracture ($\tau^\infty_f / \tau_y \leq 0.95$) and cohesive zone strength-controlled fracture ($\tau^\infty_f / \tau_y > 0.95$).  It is also instructive to plot the boundary between the regime in which failure is dictated by the elastic properties of the adhesive layer and the regime in which failure is dictated by the elastic properties of the substrate.  This condition is approximated by the geometric relation
\begin{equation} \label{eq:boundary1}
\left( l_s + a \right) <1.1 h
\end{equation}
\noindent Consequently, there are four regimes of behaviour A to D for the centre-cracked sandwich plate.  Regimes A and B satisfy the criterion (\ref{eq:boundary1}), and the fracture strength of the joint is given by (\ref{eq:shortcrack}).  In contrast, regimes C and D do not satisfy (\ref{eq:boundary1}) and the fracture strength of the joint is given by (\ref{Eq:lsFINAL2}).  The shear strength of the joint is dictated by the cohesive zone strength in regimes A and D, and by the cohesive zone toughness in regimes B and C.  Sketches are included in Fig. \ref{fig:MapFigure} to illustrate the relative magnitude of the length scales   in regimes A to D, where $R_p$ is the length of the cohesive zone at fracture, and is, in general, different from the material length scale $l_s$.

\subsubsection{Numerical verification}
\label{Sec:NumVerify}

It remains to verify the accuracy of the analytical formulae, Eqs. (\ref{eq:shortcrack}) and (\ref{Eq:lsFINAL2}), by finite element calculations of the strip-yield model.  The strip-yield model is implemented in ABAQUS/Standard by making use of cohesive elements, see Ref. \cite{EFM2017} for details. In brief, mode I opening is suppressed by the cohesive zone by the imposition of a penalty function, and only mode II sliding along the cohesive zone surfaces, of shear strength $\tau_y$, is permitted, as for the case of the semi-infinite crack in the sandwich layer.  A total of approximately 87,000 plane strain, quadratic elements have been used. The crack tip sliding displacement is determined as a function of increasing $\tau^\infty_f/\tau_y$ for selected crack lengths, for $h/W=0.01$ and $E_1/E_2=10$ and 100. Recall  that the reference length is related directly to the crack tip sliding displacement via
\begin{equation}
l_s = \frac{1}{\pi(1-\nu_2^2)}\frac{E_2 \delta_c}{\tau_y} 
\end{equation}

\begin{figure}[H]
  \makebox[\textwidth][c]{\includegraphics[width=0.8\textwidth]{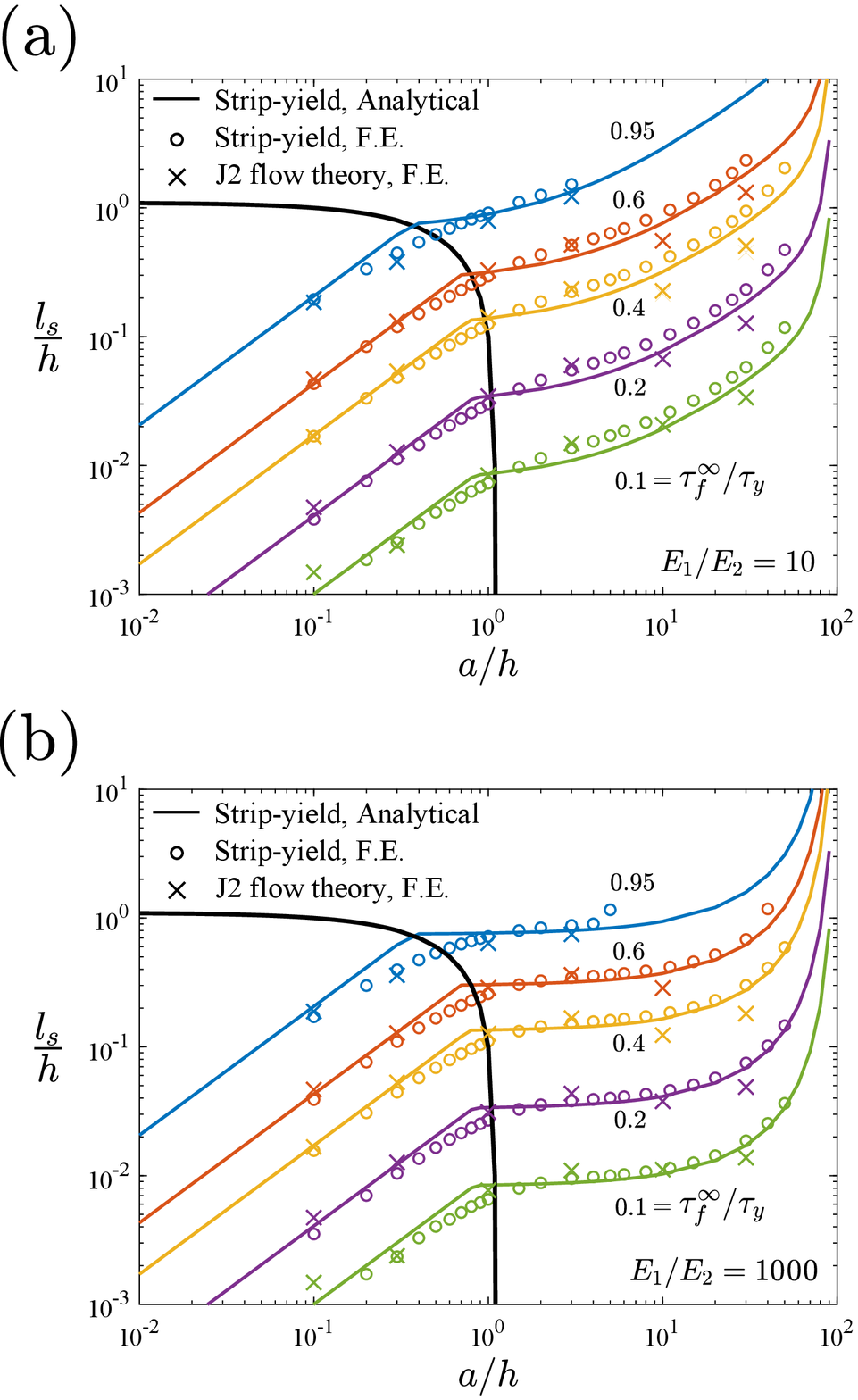}}%
  \caption{Failure map of $l_s/h$ versus normalized crack length $a/h$ for $h/W=0.01$ and mismatch (a) $E_1/E_2=10$ and (b) $E_1/E_2=1000$. The black solid line separates the regimes A and B from the regimes C and D.}
  \label{fig:Mode2Fred}
\end{figure}

\noindent The analytical predictions (for the toughness-controlled regimes B and C) are compared with the finite element predictions in Fig. \ref{fig:Mode2Fred} for $E_1/E_2=10$ and 100.  The accuracy of the analytical formulae is acceptable for the purpose of the construction of failure maps. \\

The above analysis assumes that the shear version of the strip-yield model is adequate for modelling the fracture process zone at the crack tip.  As already discussed above in the context of the semi-infinite crack in a sandwich layer, the strip-yield model also serves the purpose of an idealisation for crack tip plasticity.  Indeed, we have already concluded that the strip-yield model is accurate for this purpose for the semi-infinite crack, for which a remote K-field exists.  A similar exercise can be performed for the centre-cracked sandwich plate for which a remote K-field may, or may not, exist, depending upon the load level.  A series of finite element calculations have been performed whereby the layer is made from an elastic, ideally plastic solid that satisfies J2 flow theory in order to determine whether the strip-yield model is able to predict the crack tip sliding displacement.  The comparison of the finite element predictions for the strip-yield model and for J2 flow theory (absent a cohesive zone) is included in  Fig. \ref{fig:Mode2Fred}.  It is concluded that the strip-yield model gives accurate insight into the crack tip field for a wide range of load level and crack length.

\section{Concluding remarks}
\label{Sec:Concluding remarks}

An analytical and numerical treatment of mode II fracture of adhesive joints is reported. Two geometries are considered, a boundary layer formulation whereby a remote $K_{II}$ field is prescribed, and a centre-cracked plate subjected to remote shear stress. In both cases the adhesive layer is sandwiched between two elastic substrates and insight is gained into the role of the material mismatch on the macroscopic fracture strength.  Both elastic-brittle, and  elastic-plastic sandwich layers are considered.  New analytical solutions for determining the strength of adhesive joints are presented and are verified by detailed finite element calculations. These solutions enable the prediction of macroscopic fracture strength as a function of crack length,  height of the sandwich layer, geometry of the plate, elastic modulus mismatch and toughness of the adhesive. The main findings for a semi-infinite crack in the sandwich layer are:\\

\noindent (i) A region of constant shear stress exists ahead of the crack tip, of size that scales with the layer height and the substrate/layer  modulus ratio. The existence of this extensive zone of uniform stress compromises the existence of a remote $K$ field and hinders the use of linear elastic fracture mechanics for engineering assessment of adhesive joints.\\

\noindent (ii) The ratio between  normal and shear crack tip displacement $\delta_I/\delta_{II}$  almost vanishes beyond a distance of approximately 0.4 times the layer thickness, independently of the elastic mismatch and position of of cracking plane within the adhesive layer. This result justifies the use of a pure mode II strip-yield model. \\

Fracture maps have been constructed for a centre-cracked sandwich plate; the predictions of simple analytical formulae are in good agreement with detailed finite element calculations.  Regimes of behaviour are identified, such as the regime wherein failure is dictated by the modulus of the layer, and a regime wherein failure is dominated by the modulus of the substrate.  The sensitivity of the macroscopic shear strength of the panel to the ratio of crack length to layer height is also made quantitative.

\section{Acknowledgments}
\label{Acknowledge of funding}

The authors would like to acknowledge financial support from the European Research Council in the form of an Advance Grant (MULTILAT, 669764), and from the Interreg 2 Seas Mers Zee{\"{e}}n EU programme (QUALIFY project). E. Mart\'{\i}nez-Pa\~neda additionally acknowledges financial support from Wolfson College Cambridge (Junior Research Fellowship) and from the Royal Commission for the 1851 Exhibition through their Research Fellowship programme (RF496/2018). I.I. Cuesta wishes to thank the Department of Engineering of Cambridge University for providing hospitality during his research stay.

\processdelayedfloats % This is basically to include the figures here, before the appendix


%\appendix
%\section{Theory}

\bibliographystyle{elsarticle-num}
\bibliography{library}


\end{document}